\documentclass[aps,reprint,superscriptaddress,showkeys,amsmath,amssymb,floatfix]{revtex4-2}

\usepackage{graphicx}
\usepackage{subcaption}
\usepackage[dvipsnames]{xcolor}
\usepackage{amsfonts}
\usepackage{braket}
\usepackage{siunitx}[=v2]
\usepackage{setspace}
\usepackage{hyperref}

\renewcommand{\Im}{\textup{Im}}
\def\O{\mathcal{O}}
\DeclareMathOperator{\tr}{Tr}

\newcommand{\dext}{\text{d}}

\newcommand{\der}{d}
\newcommand{\iu}{\ensuremath{\text{i}}}

\newcommand{\GN}{G}

\newcommand{\Mco}{M}

\begin{document}

\bibliographystyle{apsrev4-2}

\title{Gravitational Pair Production and Black Hole Evaporation}

\author{Michael F. Wondrak}
\email[]{m.wondrak@astro.ru.nl}
\affiliation{Department of Astrophysics/IMAPP, Radboud University, P.O.\ Box 9010,
6500 GL Nijmegen, The Netherlands}
\affiliation{Department of Mathematics/IMAPP, Radboud University, P.O.\ Box 9010,
6500 GL Nijmegen, The Netherlands}

\author{Walter D. van Suijlekom}
\email[]{waltervs@math.ru.nl}
\affiliation{Department of Mathematics/IMAPP, Radboud University, P.O.\ Box 9010,
6500 GL Nijmegen, The Netherlands}

\author{Heino Falcke}
\email[]{h.falcke@astro.ru.nl}
\affiliation{Department of Astrophysics/IMAPP, Radboud University, P.O.\ Box 9010,
6500 GL Nijmegen, The Netherlands}

\date{13 March 2023}

\begin{abstract}%
We present a new avenue to black hole evaporation using a heat-kernel approach analogous as for the Schwinger effect. Applying this method to an uncharged massless scalar field in a Schwarzschild spacetime, we show that spacetime curvature takes a similar role as the electric field strength in the Schwinger effect. 
We interpret our results as local pair production in a gravitational field and derive a radial production profile. The resulting emission peaks near the unstable photon orbit. Comparing the particle number and energy flux to the Hawking case, we find both effects to be of similar order. However, our pair production mechanism itself does not explicitly make use of the presence of a black hole event horizon.
\end{abstract}


\maketitle

\emph{Introduction.}---%
According to quantum mechanics, a vacuum state is populated by virtual particle pairs undergoing spontaneous creation and annihilation processes. These quantum fluctuations can turn into real particle pairs in the presence of a background field. The most prominent example of such a process is the Schwinger effect predicting the creation of charged particle pairs in the presence of an electric field~\cite{Euler:1936,Schwinger:1951nm}, see Refs.~\cite{Dunne:2004nc,Dunne:2012vv} for a review. According to the standard interpretation, the particles of a spontaneously created virtual pair are accelerated in opposite directions by the external field. If the particles separate far enough within the time granted to them by the Heisenberg uncertainty principle, i.e.\ if the virtual particles gain enough energy over the distance of a Compton wavelength to obey the relativistic energy--momentum relation 
$E^2 = m^2 +\vec{p}^2$,
they become real. 
The threshold for the electric field strength to considerably create electron--positron pairs via the Schwinger effect, 
$\SI{\sim e18}{V/m}$, 
currently lies out of experimental reach for static configurations. However, pair production by strong electric fields may become accessible for instance in high-intensity laser beams \cite{Kohlfurst:2021skr,Fedotov:2022ely}, in the boosted electromagnetic field in ultraperipheral heavy-ion collisions, or in exotic atoms where an atomic electron shell is occupied by a more massive particle, e.g.\ a muon~\cite{Paul:2020cnx}. In condensed matter physics, a Schwinger-analog effect has been discovered in graphene recently~\cite{Berdyugin:2022}, see also~\cite{Katsnelson:2012tp,Katsnelson:2013zsa}.

Particle production also occurs in black hole spacetimes as predicted by Hawking~\cite{Hawking:1974rv,Hawking:1975vcx}. In a heuristic explanation, it originates from virtual particle--anti-particle pairs turning real: Close to the event horizon, one particle might fall into the black hole preventing a re-annihilation. The outside particle contributes to Hawking radiation~\cite{Hawking:1975vcx}. 
A mathematical treatment of the Hawking effect referring to quantum fluctuations is based on the tunneling probability of positive-energy modes outwards or negative-energy modes inwards through the event horizon~\cite{Parikh:1999mf}. 

In general, derivations of the Hawking effect require the existence of a future event horizon which is a global concept requiring knowledge over the whole spacetime also at late times. 
Furthermore they rely on time dependence such as induced by matter infall forming an event horizon or a coupling to an external thermal heat reservoir. 
According to Ref.~\cite{Visser:2001kq}, the global concept may be relaxed to a local, observer-dependent requirement for an apparent horizon.
Other processes of particle production in curved spacetimes have been discussed within 
cosmology (see e.g.\ \cite{Parker:1968mv,Schaefer:1979mb} and \cite{Ford:2021syk} for a review), 
small time-dependent anisotropies (\cite{Zeldovich:1971mw,Zeldovich:1977vgo,Birrell:1979pi})
or arise from self-interaction of the quantum field \cite{Lotze:1978ce,Birrell:1978zs}. 
In some of these situations, no horizon is present.

Sometimes a heuristic connection between the Schwinger and Hawking effect is drawn for illustrative purposes, e.g.\ \cite[sect.~12.8]{ShapiroTeukolsky2004} and \cite[sect.~1.7]{MukhanovWinitzki2007}. 
Previous studies aiming to show such a connection in specific cases followed the Hawking description in that they used so-called Bogoliubov coefficients to construct effective actions confined to the horizon of black holes~\cite{Stephens:1989fb,Parentani:1991tx,Kim:2011fs,Kim:2012wg} or focused on tunneling through  horizons~\cite{Srinivasan:1998ty,Parikh:1999mf}.

\begin{figure}[th]
\includegraphics[width=\linewidth]{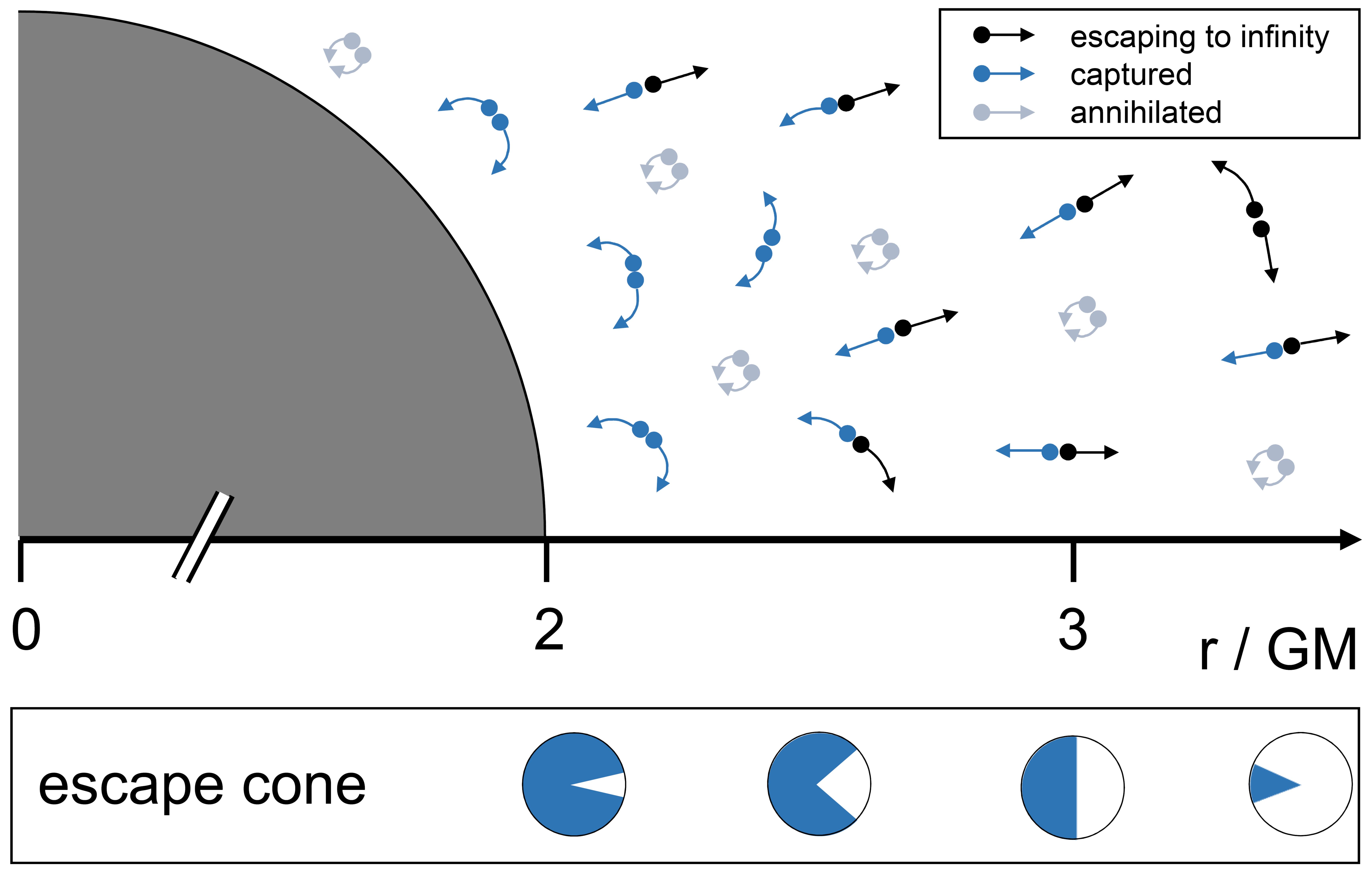}
\caption{\label{fig:schematic} Schematic of the presented gravitational particle production mechanism in a Schwarzschild spacetime. 
The particle production event rate is highest at small distances, whereas the escape probability (represented by the increasing escape cone (white)) is highest at large distances.}
\end{figure}

In this letter, we show the existence of a local gravitational particle production mechanism in curved spacetimes similar to the Schwinger effect for electric fields. 
Virtual pairs are separated by local tidal forces and become real. For black holes, some fraction of them will escape to infinity while others will be recaptured, according to the standard escape fraction for massless particles created near a black hole (see Fig.~\ref{fig:schematic}).
Both the Schwinger and our particle production effect arise as two facets from the same mathematical object: the one-loop effective action $W$ in the heat-kernel representation~%
\footnote{
We follow the $(-,+,+,+)$ metric signature convention for Lorentzian manifolds, set $c=1$, and employ Lorentz--Heaviside units.%
}.
Applied to a massless complex scalar in a homogeneous electric background field, we re-obtain the Schwinger effect, i.e.\ we demonstrate that our formalism based on the weak-field form of the heat kernel agrees with the closed-form expression. 
Applied to a massless real scalar in the Schwarzschild background gravitational field, the predicted amount of particle production is comparable to the a priori independent Hawking effect. 
Our approach applies to general mass configurations and does not explicitly invoke the presence of an event horizon other than for calculating an escape fraction.

\emph{Particle production.}---%
In the following, we investigate the stationary production of real massless scalar particles by weak electric and gravitational background fields. 
Going back to Schwinger~\cite{Schwinger:1951nm}, the associated vacuum non-persistence rate is encoded in the imaginary part of the one-loop effective action, which is defined as
\begin{align}
W_{{}_\text{E}}
&= \frac{\hbar}{2}\, \tr \ln H/\tilde{\mu}^2\\
\label{eq:W-path_integral}
\begin{split}
&= -\frac{\hbar}{2}\, \left(\frac{\tilde{\mu}}{\hbar}\right)^{2z} \int \dext^D x\, \sqrt{g_{{}_\text{E}}}\, \int_0^\infty \frac{\dext s}{s^{1-z}}\, \\
&\qquad{} \times \int_{x(0)=x(s)} \mathcal{D}x(\tau)\, \exp\!\left(-\frac{1}{\hbar}\, S_\text{e}[x(\tau)]\right)
\end{split}
\end{align}
in the Euclidean case with metric $g_{{}_\text{E} \mu\nu}$ and can be Wick rotated to Lorentzian signature in the case of a stationary spacetime. 
In the first line, the effective action is expressed in terms of the second-order elliptic differential operator for the scalar field, $H$. We apply a zeta-function regularization including a mass scale $\tilde{\mu}$ and a parameter $z \in \mathbb{C}$. The second line makes the interpretation in terms of field excitations localized at $x(\tau)$ evident, which classically would evolve according to $H$ as the Hamilton operator for a duration of the so-called proper time $s$.
In terms of the path integral, all excitation trajectories are considered which start and end in the same point, and are weighted according to the classical action~\cite{Strassler:1992zr}, 
\begin{equation}
S_\text{e}[x(\tau)]
= \int_0^\tau \dext \tau^\prime\; \left[ p_\mu \dot{x}^\mu -H/\hbar \right].
\end{equation}
In the Lorentzian counterpart, a closed trajectory requires the field excitation to move forward and backward in the exterior time. It can be interpreted as the virtual creation and annihilation of a particle--anti-particle pair~\cite{Dunne:2004nc} (represented by light gray trajectories in Fig.~\ref{fig:schematic}).

The sum of all particle--anti-particle paths, i.e.\ the path integral in eq.~\eqref{eq:W-path_integral}, is also referred to as the coincidence limit of the heat kernel. 
It can be factorized in a purely classical contribution, $\exp(-m^2 s)$, and a contribution from fluctuations for which we utilize the Barvinsky--Vilkovisky expansion~\cite{Barvinsky:1990up}.
This enables us to derive the imaginary part of the Lorentzian effective action for a massless scalar in the weak-field limit, which is described in detail in Appendix~\ref{sec:App-Derivation_Eff_Action}: 
\begin{align}
\begin{split}
\Im(W)
&= \frac{\hbar\, N}{64\pi}\, \int \dext^4 x\, \sqrt{-g}\; \bigg[ 
 \frac 1 2 \left(\xi -\frac 1 6 \right)^2 R^2\\
&\quad{} +\frac{1}{180} \left(R_{\mu\nu\rho\sigma} R^{\mu\nu\rho\sigma} 
  -R_{\mu\nu} R^{\mu\nu} \right) 
 +\frac{1}{12}\, \Omega_{\mu\nu} \Omega^{\mu\nu}
\bigg]
\end{split}
\label{eq:Intro-EffAction}
\end{align}
Here, $N$ is the number of degrees of freedom, i.e.\ $1$ for a real and $2$ for a complex scalar field, 
and $\xi$  the gravitational coupling parameter.
The Ricci scalar and tensor are denoted by $R$ and $R_{\mu\nu}$, and 
the Riemann tensor by $R^\mu_{\;\nu\rho\sigma}$.
The gauge-field curvature in electrodynamics reads 
$\Omega_{\mu\nu} = \iu q F_{\mu\nu}/\hbar$, 
which contributes only for a charged, i.e.\ complex, scalar. 
As we show below, this effective action allows a unified description of the Schwinger effect and our gravitational particle production mechanism. 

Equation~\eqref{eq:Intro-EffAction} allows us to directly calculate the Schwinger effect, i.e.\ the vacuum instability rate density with regard to the spontaneous creation of massless particles in a homogeneous electric background field $\vec{E}$. In Appendix~\ref{sec:App-Equivalence_Schwinger}, we obtain 
$2\Im (\mathcal{L}_\text{eff})
= (q \vec{E})^2/96 \pi \hbar$,
where $\mathcal{L}_\text{eff}$ is the effective Lagrange density belonging to the effective action $W$.
This result agrees with the work by Euler and Heisenberg~\cite{Euler:1936} and Schwinger~\cite{Schwinger:1951nm} on electron--positron pairs when generalized to a massless complex scalar field  (see also Ref.~\cite{Yildiz:1979vv} for a generalization to massless vector bosons in non-Abelian Yang--Mills theory, in particular QCD).

\emph{Application to the Schwarzschild spacetime.}---%
We can now apply the same method to a quantum field in a curved black hole spacetime. 
We consider the asymptotically flat, spherically symmetric, and static Schwarzschild spacetime in $(3+1)$ dimensions,
\begin{align}
\begin{split}
ds^2 
&= -\left (1- \frac{2\GN\Mco}{r} \right) \dext t^2 
  +\left (1- \frac{2\GN\Mco}{r} \right)^{-1}  \dext r^2 \\
&\quad{} + r^2 \left( \dext \theta^2 +\sin^2 \theta\, \dext \phi^2 \right),
\end{split}
\end{align}
where $\GN$ is the gravitational constant and $M$ is the mass of the black hole.
Because of Ricci flatness, $R_{\mu\nu}=0$, the only non-vanishing contribution to the imaginary part of the integrand of the effective action~\eqref{eq:Intro-EffAction}
stems from the Kretschmann scalar 
$R_{\mu\nu\rho\sigma}R^{\mu\nu\rho\sigma} 
= 48\, {(\GN \Mco)}^2/r^6$.
Upon spatial integration over the black hole exterior, which is the spacetime region causally connected to outside observers (see the escape probability factor below), 
one obtains for the effective Lagrange function
$\Im(L_\text{eff})
= \hbar/1440\, \GN M$.
This predicts a rate density of particle production events of 
\begin{equation}
\frac{\der N_\text{e}}{\der t}
= \frac{1}{\num{720}\, \GN M}
\approx \num{1.39e-3}\, {(\GN M)}^{-1},
\end{equation}
where the rate refers to subsequent slices of constant Schwarzschild time $t$, the proper time for a static observer at infinity.

Per event, an arbitrary number of particle pairs can be produced, although in the majority of events only one particle pair is created. As has been shown for the Schwinger effect, the mean number of pairs can be extracted from the closed-form expression of the heat kernel and is encoded by the first pole~\cite{Nikishov:1969tt,Lebedev:1984mei}. Thus we can define a pair-per-event correction factor 
\begin{equation}
f_\text{PPE}
= \frac{\text{(rate density of particle pairs)}}{\text{(rate density of events)}}
= \frac{12}{\pi^2}
\end{equation} 
in the bosonic Schwinger case.
Because of the use of a weak-field expansion for the heat kernel, we cannot distinguish individual pole contributions, but instead adopt the factor $f_\text{PPE}$ for the gravitational particle production mechanism.
Then we find $\sim\num{3.38e-3}\, {(\GN M)}^{-1}$ for the total number rate of particles produced. 

\begin{figure}[th]
\includegraphics[width=0.48\textwidth]{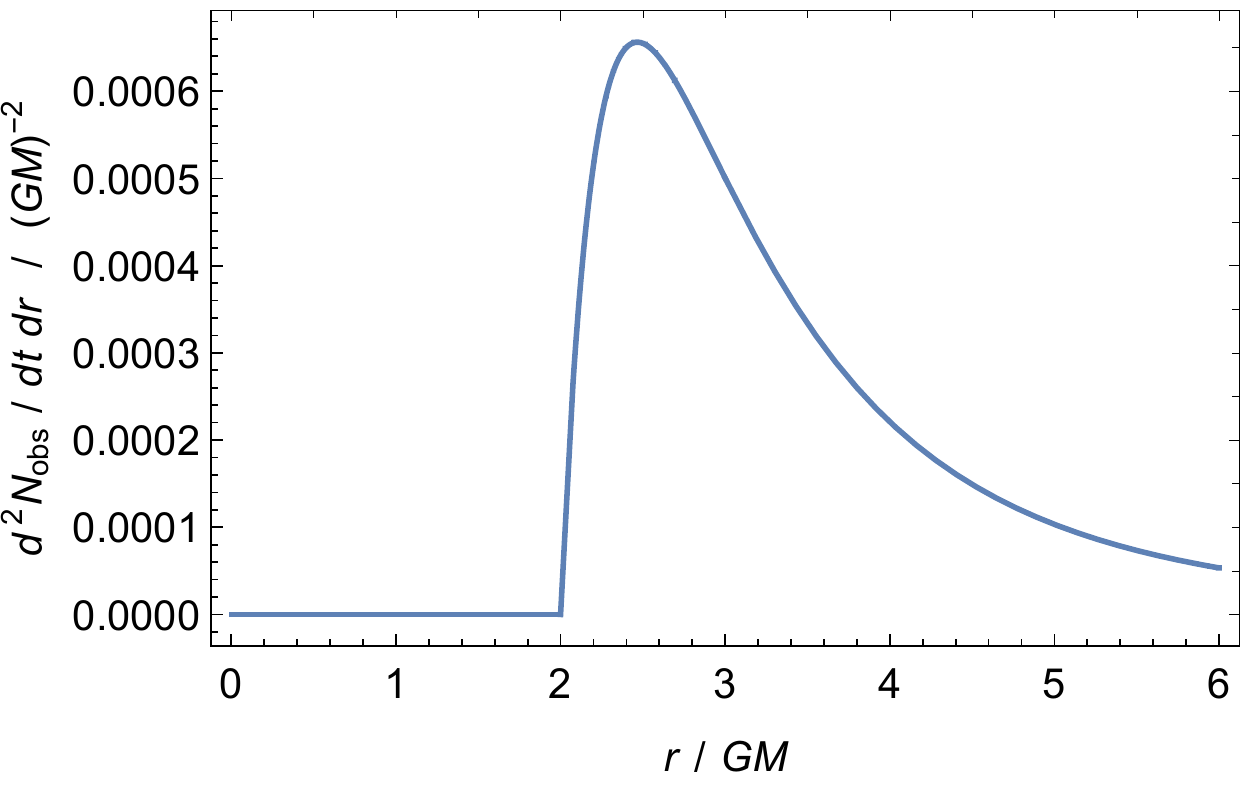}
\caption{\label{fig:dNobsdtdr} Radial profile of the production of particles escaping to infinity per unit Schwarzschild time $t$.}
\end{figure}

If produced at a larger distance, particles are less likely to be captured by the black hole (blue trajectories in Fig.~\ref{fig:schematic}) than to escape to infinity (black).
The capture cone opening angle $\alpha_\text{capt}$ for lightlike geodesics as measured by a local static Schwarzschild observer is given by~\cite{Synge:1966okc}
\begin{equation}
\sin^2 \alpha_\text{capt}
= \frac{27\, (\GN \Mco)^2\, (1 -2 \GN\Mco/r) }{r^2}
\end{equation}
and sketched in blue in the lower panel of Fig.~\ref{fig:schematic}.
Under the assumption of isotropic emission for such an observer, we define the escape probability $f_\text{esc}
= \left(1 +\cos \alpha_\text{capt} \right)/2$.
Taking these factors into account, Fig.~\ref{fig:dNobsdtdr} displays the radial distribution of the origin of particles which reach infinity per unit step in Schwarzschild time $t$. 
The distribution is non-zero throughout the black hole exterior featuring a steep drop towards the Schwarzschild radius and a power-law decrease at large distances.
The highest production rate density for escaping particles occurs around $2.32\, \GN M$ while the highest number of particles per spherical shell is produced at approx.~$2.47\, \GN M$, which is roughly half way between the event horizon and the circular photon orbit at $3\, \GN M$.
The total particle rate observed at infinity is given by
\begin{equation}
\frac{\der N_\text{obs}}{\der t}
= \frac{\num{2059}}{\num{170100}\, \pi^2\, \GN M}
\approx \num{12.3e-4}\, {(\GN M)}^{-1}.
\end{equation}

In order to estimate the energy flux of the created particles at infinity, we infer the energy of a gravitationally produced particle from comparison with the Schwinger effect in two steps. First, we deduce the characteristic amount of energy which a virtual particle gains during the Schwinger production process. It depends on the electric field strength and amounts to 
$E_\text{Schwinger} 
= (\hbar^2\, q^2 \vec{E}^2)^{1/4}
= \hbar\, (\Omega_{\mu\nu} \Omega^{\mu\nu} / 2)^{1/4}$ as obtained from the exponential term present in the closed-form expression of 
$\Im\!\left(\mathcal{L}_\text{eff}\right)$, see eq.~\eqref{eq:Im-Euler_Heisenberg} in Appendix~\ref{sec:App-Equivalence_Schwinger}. 
Second, we derive a correspondence between electric and gravitational field strengths based on requiring the same rate of particle production events. 
According to eq.~\eqref{eq:Intro-EffAction}, 
the gauge-field expression 
$\Omega_{\mu\nu} \Omega^{\mu\nu} / 12$
leads to the same imaginary part of the effective action as the gravitational expression
$R_{\mu\nu\rho\sigma} R^{\mu\nu\rho\sigma} / 180$
as it reads for Ricci-flat spacetimes.
Thereby we obtain the formal replacement rule 
$\Omega_{\mu\nu} \Omega^{\mu\nu}
\to 2/30 \times R_{\mu\nu\rho\sigma} R^{\mu\nu\rho\sigma}$, 
such that the characteristic energy scale for the gravitational particle production reads 
\begin{equation}
E_\text{curv}
= \hbar\, {\left(R_{\mu\nu\rho\sigma}R^{\mu\nu\rho\sigma}/30\right)}^{1/4}.     
\end{equation}
In general, one can expect the characteristic energy scale to carry most weight in the spectrum because lower-energy particles have wavelengths too large to resolve the spacetime curvature, and higher-energy particles perceive the spacetime as flat because of the equivalence principle. 
Taking into account the gravitational redshift, we obtain for the energy flux
\begin{equation}
\label{eq:EFlux_SSCircumference}
\frac{\der E}{\der t}
\approx \num{12.8e-5}\, \hbar\,{(\GN M)}^{-2}.
\end{equation}

The characteristic energy scale can be represented in a more intuitive way by associating it with an effective temperature scale which we would obtain if the spectrum was thermal.
By matching to $E_\text{curv}$ the average energy of particles in blackbody radiation, 
$\left\langle E \right\rangle = \pi^4/30\zeta(3) \times k_\text{B} T$, 
we find for a static observer located at radial position~$r$ 
\begin{align}
T_\text{eff}(r)
&= \frac{30\zeta(3)}{\pi^4}\, \frac{E_\text{curv}(r)}{k_\text{B}}\\
&\approx \SI{164}{nK}\, {\left(\frac{r}{2.5\, \GN M}\right)}^{-3/2}\, {\left(\frac{M}{M_\odot}\right)}^{-1},
\end{align}
with the Riemann zeta function $\zeta(x)$ and the solar mass $M_\odot \approx \SI{1.988e30}{kg}$. 

In order to provide some context for the results from our local gravitational particle production mechanism, we contrast them with the properties of Hawking radiation. 
Taking into account the gravitational redshift effect, an observer at infinity finds thermal radiation at the Hawking temperature
\begin{align}
T_\text{H}
= \frac{1}{8\pi}\, \frac{\hbar}{k_\text{B} \GN M}
\approx \SI{62}{nK}\, {\left(\frac{M}{M_\odot}\right)}^{-1}.
\end{align}
Around the distance of highest particle production, $r \approx 2.5\, \GN M$, a stationary observer would measure the Hawking radiation to have the local (Tolman) temperature of
$T_\text{H}/\sqrt{-g_{00}} \approx \SI{138}{nK}$, which is exceeded by the effective temperature from our effect by a factor $\sim (1.9)^{1/4}$.
According to the Stefan--Boltzmann law, this implies an increased particle production rate by a factor of around $1.9$. Independently, our predictions for the particle and energy flux actually also lie within a factor $\sim 1.9$ above the Hawking fluxes, see Appendix~\ref{sec:App-Hawking_predictions}. 

In contrast to radiation in the Hawking case, which is in a thermal mixed state, the radiation in our case also includes correlated pairs when the emission direction of both the particle and anti-particle fall into the escape cone. This is similar to the Schwinger effect, where the produced particles do not follow a thermal spectrum~\cite{Lebedev:1984mei}.

We point out the favorable property that our results for the total particle number and energy flux are expressed only in terms of curvature invariants and thus are invariant under coordinate transformations. 
However, they depend on the radial profile of the escape probability, for which we made two assumptions. First, we adopted a measure in the geometrical-optics approximation, i.e.\ the solid-angular size of the escape cone. This is a rather reliable assumption given the $6\%$ accuracy in the DeWitt approximation for the graybody factor discussed in Appendix~\ref{sec:App-Hawking_predictions}. 
Second, we made the assumption that the emission of real particle--anti-particle pairs is isotropic for a static observer. 
If we assumed an isotropic emission in the frame of a freely infalling Painlev\'e observer, the distribution looks beamed towards the black hole for a static observer and the particle flux would be reduced to $\sim 56\%$ of the Hawking value. 
Actually one might expect an anisotropic emission with a stronger radial component because of the anisotropic spacetime curvature. 

\emph{Discussion.}---%
In this letter we have shown that the formalism underlying the Schwinger effect can be applied to curved spacetimes, unifying the description of the decay of electric and gravitational fields in the presence of a scalar quantum field. In our non-perturbative path-integral approach, particle production arises from the separation of virtual field excitations and is encoded in the imaginary part of the one-loop effective action. 
The effective action is defined by the trace of the propagation operator which implies that the corresponding path integral in eq.~\eqref{eq:W-path_integral} is performed only over paths with coincident start and end point.
Such an event can be regarded as the origin of a virtual particle--anti-particle pair, and the associated path sections as the corresponding particle trajectories. 

Virtual particles can exist within a portion of spacetime that is constrained by the Heisenberg uncertainty principle.
The weighted path densities around classical trajectories indicate the probability distribution of a particle location, which one could call a ``Heisenberg uncertainty cloud.'' In flat spacetime, the volume of these clouds is identical for a particle and anti-particle, leading to complete annihilation. In curved spacetimes, these clouds are not necessarily identical, as particle trajectories pass through different portions of spacetime, which suppresses the particles' coherence, impacts their annihilation probability and leads to the appearance of real pairs. 
Such particle decoherence effects occurring in curved spacetimes may even be experimentally testable~\cite{Pikovski:2013qwa}. 

Alternatively, the non-overlapping portions of the uncertainty clouds between particle and anti-particle could also be interpreted as being separated by an apparent horizon, reminiscent of the horizon encountered by an accelerating observer seeing Unruh radiation~\cite{Unruh:1976db}.

In a Ricci-flat spacetime, such as in asymptotically flat black-hole spacetimes and in particular the Schwarzschild solution, the integrand given in eq.~\eqref{eq:Intro-EffAction} reduces to the square of the Weyl tensor $C_{\mu\nu\rho\sigma}$, i.e.\ 
$R_{\alpha\beta\gamma\delta} R^{\alpha\beta\gamma\delta}
= C_{\alpha\beta\gamma\delta} C^{\alpha\beta\gamma\delta}$.
Because the Weyl tensor indicates volume-preserving curvature components, it particularly encapsulates tidal forces. Hence, while in the Schwinger case the electric field separates virtual pairs, this role is taken on by tidal forces in gravitational fields, which act on all types of particles (including photons) -- not just charged ones. 
The energy of the pairs produced ultimately has to come from the gravitational field via its coupling to the corresponding quantum field. 

In analogy to the Schwinger effect in electric background fields, we use heat-kernel techniques for describing gravitational pair production. 
The action for the fictitious field excitation, $S_\text{e}$, factorizes in a classical and a fluctuating part. In the Schwinger case, this can be evaluated exactly and allows one to derive an expression of particle production for massive and massless particles.
In contrast, no closed-form solution for the fluctuating part is known to date in a gravitational background (see~\cite{Bekenstein:1981xe} for a quadratic approximation). 
However, an expansion in weak fields is known, see e.g.\ \cite{Barvinsky:1990up,Codello:2012kq}, which enables us to treat the massless case. 

Massless particles are of special interest because they dominate the black hole emission for most of the black hole's lifetime. Only at the end of black hole evaporation (below the mass scale of $\SI{e14}{kg} \sim \SI{e-16}\, M_\odot$, corresponding to subatomic Schwarzschild radii below $\SI{e-12}{m}$) do massive particles contribute a relevant share to the radiation~\cite{Page:1976df} before potential extra dimensions become relevant~\cite{Hossenfelder:2001dn}.
Moreover, the scalar field considered here is a generic model for one massless bosonic degree of freedom, e.g.\ one polarization state of the photon (conformal coupling~\cite{Cote:2019kbg}, $\xi=1/6$) or of the graviton (minimal coupling~\cite{Lifshitz:1945du}, $\xi=0$).
Interestingly, our results show that the production rate is independent of the gravitational coupling constant $\xi$ in Ricci-flat spacetimes.

The particle production effect which we have presented predicts within a factor $\sim 2$ the same fluxes, effective temperature, and their respective scalings with the black hole mass as in the Hawking case, assuming thermal emission. 
It is not immediately clear whether this is an additional effect to Hawking radiation or a generalization thereof. At least, for the radiation escaping between $2 \GN M < r \leq 3 \GN M$, the other half of the pair will fall into the event horizon, similar to what is described in the popular explanation of Hawking radiation. 
However, it is a local approach. The presence of a black hole event horizon does not enter our derivation, apart from the escape factor. 
This is in contrast to Hawking radiation, where the presence of a global event horizon or time dependence is considered vital. Among these, the approaches by Parikh and Wilczek~\cite{Parikh:1999mf} and Mukhanov, Wipf, and Zelnikov~\cite{Mukhanov:1994ax} have aspects in common with our work.
However, both these approaches inherently require the reduction to $1+1$ dimensions and thus cannot be sensitive to local anisotropies and tidal forces. 

The popular description of Hawking radiation suggests that most of the particle production happens at an infinitesimal distance from the event horizon. For our mechanism, we explicitly deduce a radial production profile for evaporating black holes. Its peak is determined by the counteraction of the rate of local pair creation and the escape probability, see Fig.~\ref{fig:schematic}.
The local, generally covariant expression for the imaginary part of the effective action in eq.~\eqref{eq:Intro-EffAction} actually indicates that this gravitational particle production is independent of the choice of the vacuum state of the quantum field. While there might also be state-dependent contributions encoded in the subleading higher-order behavior of the form factors, those would in general be tied to non-local terms~\cite[pp.~76,~178]{BirrellDavies1984}. 

Similarly to Ref.~\cite{Parikh:1999mf}, we also expect our particle production mechanism to backreact on the spacetime. However, the spacetime background we have used here is constant by construction which implicitly requires an external energy reservoir, the thermal nature of which can be debated as in the case of the Unruh effect~\cite{Buchholz:2014jta,Buchholz:2015fqa}. 
Clearly, if there is radiation without an external energy reservoir, it means that the spacetime is non-stationary. In consequence, this would imply that the black hole evaporates indeed. In principle, the pair production process should also continue inside the event horizon, but this would not be observable outside. The process would diverge at $r=0$ and the associated backreaction could potentially tame the curvature singularity in the center. The weak-field assumptions of our work, however, would not extend to this extrapolation.

Taken at face value, our local approach could imply that also mass configurations without a global event horizon would radiate and eventually decay. Furthermore, evaporation processes might not necessarily lead to an information paradox. For approaching these questions, the entropy of the central object and thermal nature of the emission are topics of further investigation.

\begin{acknowledgments}
The idea to investigate the Schwinger formalism emerged initially in a discussion with G.~Sch\"afer in the Bierkeller of the DPG Physikzentrum in Bad Honnef in July 2017 over a glass of water. The authors also thank M.~Bleicher, J.~de~Boer, G.~'t~Hooft, M.~Kaminski, M.~Niedermaier, I.A.~Reyes, S.~Vandoren, and M.~Visser for inspiring and helpful discussions. We are indebted to C.~Beenakker, K.~Landsman, and M.~Katsnelson for commenting on the manuscript.
F.~Schmidt-Kaler pointed out the relation to experiments of particle coherence in a gravitational field.
We thank the referee for very insightful comments helping us to sharpen a number of key points.
All authors contributed equally to this work.
It was supported by an Excellence Fellowship from Radboud University and a grant from NWO NWA 6201348. 
\end{acknowledgments}

\appendix
\clearpage
\onecolumngrid

\begin{center}
\textbf{\large --- Supplemental Material ---\\ $~$ \\
Gravitational Pair Production and Black Hole Evaporation}\\
\medskip
\text{Michael F.\ Wondrak, Walter D.\ van Suijlekom, and Heino Falcke}
\end{center}
\setcounter{equation}{0}
\makeatletter
\renewcommand{\thesection}{S.\arabic{section}}
\renewcommand{\theequation}{S.\arabic{equation}}

\section{Derivation of the Imaginary Part of the Massless Effective Action}
\label{sec:App-Derivation_Eff_Action}
Using heat kernel methods, we derive the imaginary part of the effective action for a massless scalar field in a gravitational background. This is closely related to the works \cite{Parker:1999td,ParkerToms2009} and references therein. 
However, in an attempt to be as rigorous as possible, we give a derivation in Euclidean signature, i.e.\ on a Riemannian manifold, using an exact expression for the heat trace obtained in covariant perturbation theory.

Consider the Euclidean action functional for a free real scalar field $\phi$ in $D$ (even) dimensions,
\begin{align}
S_{{}_\text{E}}[\phi,g_{{}_\text{E}}]
&= \frac{1}{2} \int \dext^D x\, \sqrt{g_{{}_\text{E}}}\, \left( 
 \hbar^2\, \partial_\mu \phi\, \partial^\mu \phi 
 +m^2 \phi^2
 +\hbar^2\, \xi R\, \phi^2
\right) \\
&= \frac{1}{2} \int \dext^D x\, \sqrt{g_{{}_\text{E}}}\; \phi H \phi,
\end{align}
where $H = \hbar^2 \Delta +m^2 +\hbar^2\, \xi R$, with 
$\Delta = -\nabla_\mu \nabla^\mu$, 
is a second-order elliptic differential operator. 
The mass is denoted by $m$ and the gravitational coupling parameter by $\xi$. For the effective action in a gravitational background with Euclidean metric $g_{{}_\text{E}\,\mu\nu}$ and Ricci scalar $R$, we integrate out the scalar field (so only treating the free scalar field as a quantum field) and obtain the Euclidean 1-loop effective action 
\begin{align}
W_{{}_\text{E}}[g_{{}_\text{E}}] 
&=-\hbar \ln \int \mathcal{D}\phi\, \exp\!\left(-\frac{1}{\hbar}\, S_{{}_\text{E}}[\phi,g_{{}_\text{E}}]\right) \\
&=\frac{\hbar}{2} \ln \det H / \tilde{\mu}^2\\
&=\frac{\hbar}{2} \tr \ln H / \tilde{\mu}^2\\
\label{eq:Eff_action_Heat_kernel_representation}
&=-\frac{\hbar}{2}\, \left(\frac{\tilde{\mu}}{\hbar}\right)^{2z} 
 \int_0^\infty \frac{\dext s}{s^{1-z}}\, \tr e^{-sH/\hbar^2},
\end{align}
where we used the $\zeta$-function regularization with a complex parameter $z$ and 
an arbitrary renormalization mass scale $\tilde{\mu}$ 
to keep proper physical dimensions. (See~\cite{Haw77,Vas03} for a more extensive treatment and \cite{Parker:1999td,Ohta:2020bsc} for different regularization schemes.)
Note that the heat kernel $\tr e^{-sH/\hbar^2}$ is well-defined precisely because we are working in Euclidean signature. 

\subsection{Covariant Perturbation Theory}
In \cite{Barvinsky:1990up}, Barvinsky and Vilkovisky derived an expansion of the heat trace $\tr e^{-s( \Delta+ E)}$ to any (finite) order in the curvature that is exact in the spectral parameter $s$ and in $\Delta$, which was also derived more recently by alternative methods in \cite{Codello:2012kq}. Up to second order in $R$, $E$, and the curvatures $R_{\mu\nu}$ and $\Omega_{\mu\nu}$ associated to $\Delta$, this reads (in dimension $D=4$)
\begin{align}
\label{eq:HeatKernel-BV_expansion}
\begin{split}
\tr e^{-s( \Delta+ E)}
&= \frac 1{(4\pi s)^2} \int \dext^4 x\, \sqrt{g_{{}_\text{E}}}\; \text{tr}\, \left[ 
 1 -sP +s^2 \bigl( 
  R_{\mu\nu} f_1(s \Delta) R^{\mu\nu} + R f_2(s \Delta) R 
\right.\\
&\qquad 
\left.{} 
  +P f_3(s \Delta) R +P f_4(s \Delta) P +\Omega_{\mu\nu} f_5(s \Delta) \Omega^{\mu\nu}
\bigr) \right] + \dots 
\end{split}
\end{align}
with the identification $P \equiv E -\tfrac 16 R$ and the form factors 
\begin{align}
\begin{split}
& f_1(y)
= \frac{h(y) -1 +\tfrac 1 6\, y}{y^2}, \qquad 
f_2(y)
= \tfrac 1 {288}\, h(y) -\tfrac 1 {12}\, f_5(y) - \tfrac 1 8\, f_1(y),\\
&f_3(y)
= \tfrac 1 {12}\, h(y) -f_5(y), \qquad 
f_4(y)
= \tfrac 1 2\, h(y), \qquad 
f_5(y)
= -\frac{h(y)-1}{2y},
\end{split}
\end{align}
which are expressed in terms of
\begin{equation}
h(y)
:= \int_0^1 \dext\alpha\; e^{-\alpha (1 -\alpha)\, y}.
\end{equation}
For a recent review on form factors, see~\cite{Knorr:2022dsx}.
For later use, we record the behavior of the form factor $f_1$ for small arguments,
\begin{align}
f_1 (y) 
= \frac 1 {60} -\frac y {840} +\O(y^2).
\end{align}

This expansion is determined up to topological terms which are proportional to the Pontryagin class
\begin{align}
\label{eq:pontryagin}
R^*R^* 
= R^2 -4R_{\mu\nu} R^{\mu\nu} +R_{\mu\nu\rho\sigma} R^{\mu\nu\rho\sigma} .
\end{align}
Similar to Ref.~\cite{El-Menoufi:2015cqw}, we restore these terms by comparing the leading behavior of the form factors at small $s$ to the Seeley--DeWitt heat kernel expansion, see e.g.\ Refs.~\cite[Sect.~4.1]{Gilkey1995} and \cite[Sect.~8.3]{vanSuijlekom2015}. We may rewrite the expansion~\eqref{eq:HeatKernel-BV_expansion} as
\begin{align}
\begin{split}
\tr e^{-s (\Delta +E)}
&\simeq \frac{1}{(4\pi s)^2} \int \dext^4 x\, \sqrt{g_{{}_\text{E}}}\; \text{tr}\, \left[ 
 1 -sP +s^2 \bigl( 
  R_{\mu\nu\rho\sigma} \tilde{f}_1(s \Delta) R^{\mu\nu\rho\sigma} 
  -R_{\mu\nu} \tilde{f}_1(s \Delta) R^{\mu\nu}
 \right.\\
&\qquad 
 \left.{} 
  +R \tilde{f}_2(s \Delta)R +P f_3(s \Delta) R +P f_4(s \Delta) P 
  +\Omega_{\mu\nu} f_5(s \Delta) \Omega^{\mu\nu}
\bigr) \right] + \dots
\end{split}
\end{align}
where we have defined
\begin{equation}
\tilde{f}_1 
= \frac 1 3\, f_1, \qquad 
\tilde{f}_2 
= \frac 1 3\, f_1 + f_2.
\end{equation}

\subsection{Effective Action for a Scalar Field}
For a real scalar field, we have 
$P = (\xi -\frac 1 6) R$ 
and the curvature $\Omega_{\mu\nu}$ of the gauge connection vanishes. We nevertheless keep the curvature term for generalizing the expression to a complex scalar field later. The above expression for the heat trace thus reduces to
\begin{align}
\begin{split}
\tr e^{-s (\Delta +E)}
&\simeq \frac{1}{(4\pi s)^2} \int \dext^4 x\, \sqrt{g_{{}_\text{E}}}\; \bigg[ 
 1 -s \left(\xi -\frac 1 6\right) R 
\\
&\qquad 
 {} +s^2 \left( 
  R_{\mu\nu\rho\sigma} \tilde{f}_1(s \Delta) R^{\mu\nu\rho\sigma} 
  -R_{\mu\nu} \tilde{f}_1(s \Delta) R^{\mu\nu}
  +R  f_R(s \Delta) R
  +\Omega_{\mu\nu} f_5(s \Delta) \Omega^{\mu\nu}
\right) \bigg] + \dots
\end{split}
\label{eq:heat-trace-C2}
\end{align}
with the definition
\begin{equation}
f_R 
= \tilde{f}_2 +\left(\xi -\frac 1 6\right) f_3 +\left(\xi -\frac 1 6\right)^2 f_4.
\end{equation}
Note that $f_R(y) = \frac 1 2 (\xi -\frac 1 6)^2 + \O(y)$. 
Since the classical contribution to the heat kernel reads $e^{-s m^2}$, we insert the above expression \eqref{eq:heat-trace-C2} in the effective action as follows:
\begin{align}
W_{{}_\text{E}} 
&=-\frac{\hbar}{2}\, \left(\frac{\tilde{\mu}}{\hbar}\right)^{2z} 
 \int_0^\infty \frac{\dext s}{s^{1-z}}\, 
 \tr \left(e^{-s (\Delta +\xi R)} \right) e^{-s (m^2 -\iu \epsilon)/\hbar^2}
\end{align}
We introduced a regulator $\iu \epsilon$ in order to eventually take the limit $m \to 0$.

If we are interested in the lowest order in $R_{\mu\nu\rho\sigma}$, $R_{\mu\nu}$ and $R$ and ignore derivatives thereof as those are associated by higher powers in $s$, we obtain
\begin{align}
\begin{split}
\tr e^{-s (\Delta +E)}
&\simeq \frac{1}{(4\pi s)^2} \int \dext^4 x\, \sqrt{g_{{}_\text{E}}}\; \bigg[ 
 1 -s \left(\xi -\frac 1 6\right) R 
\\
&\qquad 
 {} +s^2 \left( 
  \frac{1}{180} \left(R_{\mu\nu\rho\sigma} R^{\mu\nu\rho\sigma} 
  -R_{\mu\nu} R^{\mu\nu} \right) 
  +\frac{1}{2} \left(\xi -\frac 1 6\right)^2 R^2
  +\frac{1}{12}\, \Omega_{\mu\nu} \Omega^{\mu\nu}
\right) \bigg] + \dots
\end{split}
\label{eq:heat-trace-C2-weak}
\end{align}
Integrating eq.~\eqref{eq:heat-trace-C2-weak} against $e^{-s (m^2 -\iu \epsilon)/\hbar^2}$, while using the fact that
\begin{align}
\int_0^\infty \frac{\dext s}{s^{1-z}}\, 
 s^{j}\,  e^{-s (m^2 -\iu \epsilon)/\hbar^2}
= \Gamma(j+z)\, \left(\frac{m^2}{\hbar^2} -\iu \epsilon\right)^{-j-z} 
\end{align}
for $j = -2$, $-1$, $0$, $\ldots$ and appropriate values of $z$, we get
\begin{align}
\begin{split}
W_{{}_\text{E}}
&= -\frac{\hbar}{32\pi^2}\, \left(\frac{\tilde{\mu}}{\hbar}\right)^{2z} 
\int \dext^4 x\, \sqrt{g_{{}_\text{E}}}\; \bigg[ 
 \Gamma(-2+z) \left(\frac{m^2}{\hbar^2} -\iu \epsilon\right)^{2-z} 
 -\Gamma(-1+z) \left(\frac{m^2}{\hbar^2} -\iu \epsilon\right)^{1-z} \left(\xi -\frac 1 6 \right)R
\\
&\qquad 
 {} +\Gamma(z) \left(\frac{m^2}{\hbar^2} -\iu \epsilon\right)^{-z} \left( 
  \frac{1}{180} \left(R_{\mu\nu\rho\sigma} R^{\mu\nu\rho\sigma} 
  -R_{\mu\nu} R^{\mu\nu} \right) 
  +\frac{1}{2} \left(\xi -\frac 1 6\right)^2 R^2
  +\frac{1}{12}\, \Omega_{\mu\nu} \Omega^{\mu\nu}
\right) \bigg] + \dots
\end{split}
\end{align}

\subsection{Massless Particle Production in Weak Electric and Gravitational Fields}
In Euclidean signature, the rate of scalar particle production events is determined by the imaginary part of the effective action, $-2\Im(W_{{}_\text{E}})$. The imaginary part originates from the branch cut of the logarithm (chosen along the negative real axis) implicitly present in the expression 
$\left(m^2/\tilde{\mu}^2 -\iu \epsilon\right)^{-z}$. 
In fact, we have
\begin{align}
\left(\frac{\tilde{\mu}}{\hbar}\right)^{2z}\, \left(\frac{m^2}{\hbar^2} -\iu \epsilon\right)^{-z}
&= 1 -z\, \ln\!\left(\frac{m^2}{\tilde{\mu}^2} -\iu \epsilon\right) 
+\O(z^2),
\end{align}
where
\begin{align}
\ln\!\left(\frac{m^2}{\tilde{\mu}^2} -\iu \epsilon\right)
&= \ln \left|\frac{m^2}{\tilde{\mu}^2} -\iu \epsilon\right| 
 + \iu\, \textup{Arg}\!\left(\frac{m^2}{\tilde{\mu}^2} -\iu \epsilon\right) \\
\label{eq:App-Log_branchcut}
&= \ln \left|\frac{m^2}{\tilde{\mu}^2} -\iu \epsilon\right| 
 -\iu \pi\, \theta(-m^2)
\end{align}
in terms of the Heaviside step function $\theta$,
\begin{equation}
\theta(x) 
= \left \{ 
 \begin{matrix} 0 & x<0\\
  \frac 12 & x=0\\
  1 & x>0 
 \end{matrix} 
\right. .
\end{equation}
The value $\theta = 1/2$ in the origin is justified by the fact that 
$\textup{Arg}(-\iu \epsilon) = -\pi/2$ for $m^2 = 0$. 

Combining this with the Laurent series expansion of the $\Gamma$-functions, 
\begin{align}
\Gamma(-2+z) 
&=\frac{1}{2z} +\frac{3}{4} -\frac{\gamma_\text{E}}{2} +\O(z),\\
\Gamma(-1+z) 
&=-\frac{1}{z} -1 +\gamma_\text{E} +\O(z),\\
\Gamma(z) 
&= \frac{1}{z} -\gamma_\text{E} +\O(z),
\end{align}
we may collect only terms that are constant in $z$ to find the imaginary part of the massless renormalized effective action,
\begin{align}
\Im(W_{{}_\text{E}})
&= -\frac{\hbar}{64\pi}\, \int \dext^4 x\, \sqrt{g_{{}_\text{E}}}\; \bigg[ 
 \frac 1 2 \left(\xi -\frac 1 6 \right)^2 R^2
 +\frac{1}{180} \left(R_{\mu\nu\rho\sigma} R^{\mu\nu\rho\sigma} 
  -R_{\mu\nu} R^{\mu\nu} \right) 
 +\frac{1}{12}\, \Omega_{\mu\nu} \Omega^{\mu\nu}
\bigg] + \dots
\label{eq:ImW-exact}
\end{align}
where the dots stand for higher curvature terms.
A Wick rotation to Lorentzian spacetimes induces a sign flip for the effective action. 

We can immediately generalize the result to complex scalar fields. As a complex scalar couples to the electromagnetic 4-potential, it has a non-trivial curvature $\Omega_{\mu\nu}$ of the gauge connection. Taking into account that the complex scalar incorporates two real degrees of freedom, $N=2$, eq.~\eqref{eq:ImW-exact} only receives an additional factor of $2$.

\section{Equivalence of Approaches for the Schwinger Effect} 
\label{sec:App-Equivalence_Schwinger}
\setcounter{equation}{27} 
In this appendix, we investigate the Schwinger effect as a reference case. 
A full treatment of the Schwinger effect~\cite{Schwinger:1951nm,Dunne:2004nc} for a free complex scalar field $\phi$ with mass $m$ and charge $q$ in $D=4$ dimensions in the presence of a homogeneous electric field $\vec{E}$ using a closed-form expression for the heat kernel yields the following imaginary part of the effective Lagrange density $\mathcal{L}_\text{eff}$, 
\begin{equation}
\Im\!\left(\mathcal{L}_\text{eff}\right)
= \frac{{(q \vec{E})}^2}{16\pi^3\hbar}\, 
 \sum_{n=1}^\infty \frac{{(-1)}^{n+1}}{n^2}\, \exp\!\left(-\frac{n \pi m^2}{\hbar\, \sqrt{q^2 \vec{E}^2}}\right).
\label{eq:Im-Euler_Heisenberg}
\end{equation}
In the limit $m \to 0$, it reduces to
\begin{equation}
\Im\!\left(\mathcal{L}_\text{eff}\right)
= \frac{{(q \vec{E})}^2}{192\pi\hbar}.
\label{eq:App-Ref_Schwinger_massless}
\end{equation}

Alternatively, this expression can be readily obtained from eq.~\eqref{eq:ImW-exact} in Euclidean flat space, 
keeping in mind that a complex scalar field carries two real degrees of freedom, $N=2$,
\begin{align}
\Im(W_{{}_\text{E}})
&= -\frac{2\hbar}{64\pi}\, \int \dext^4 x\, \sqrt{g_{{}_\text{E}}}\; 
 \frac{1}{12}\, \Omega_{\mu\nu} \Omega^{\mu\nu} + \dots\, .
\end{align}
Only the curvature of the electrodynamic connection contributes, which is defined by
$\Omega_{\mu\nu} 
= \left[ \nabla_\mu,\nabla_\nu \right]
= \iu q F_{\mu\nu}/\hbar$ 
expressed in terms of $\nabla_\mu$, which reduces to a gauge-covariant derivative,
$\nabla_\mu = \partial_\mu +\iu q A_\mu/\hbar$,
and the Faraday tensor $F_{\mu\nu}$.
We consistently obtain 
\begin{align}
\Im\!\left(\mathcal{L}_{\text{eff}\,{}_\text{E}}\right)
= -\frac{2\hbar}{64\pi}\, 
 \left(-\frac{q^2 F_{\mu\nu}F^{\mu\nu}}{12 \hbar^2}\right)
= -\frac{{(q \vec{E})}^2}{192\pi \hbar}
\end{align}
up to the sign from the Euclidean derivation.

\section{Particle and Energy Fluxes for Hawking Radiation}
\label{sec:App-Hawking_predictions}
\setcounter{equation}{31}
For contextualizing our results about the gravitationally produced particle and energy flux, this appendix briefly provides the corresponding expressions for Hawking radiation~\cite{Hawking:1975vcx,Page:1976df}, 
\begin{align}
\frac{\der N}{\der t}
&= \sum_l \int_0^\infty \frac{\dext \omega}{2\pi}\; 
  N_l\, {\left|A_l\right|}^2\, \frac{1}{\exp(8\pi \GN M\, \omega) -1}
\label{eq:ParticleNumber-FiducialValue}
= \frac{27 \zeta(3)}{512 \pi^4\, \GN M}
\approx \num{6.51e-4}\, {(\GN M)}^{-1},\\
\frac{\der E}{\der t}
&= \sum_l \int_0^\infty \frac{\dext \omega}{2\pi}\; 
  N_l\, {\left|A_l\right|}^2\, \frac{\hbar \omega}{\exp(8\pi \GN M\, \omega) -1}
= \frac{9 \hbar}{\num{40960} \pi\, {(\GN M)}^2}
\approx \num{6.99e-5}\, \hbar\,{(\GN M)}^{-2}.
\end{align}
Here, the index $l$ denotes the order of the spherical harmonics of the radiation field. The degeneracy factor $N_l = 2l +1$ and the absorption probability ${\left|A_l\right|}^2$, also referred to as graybody factor, are related to the total absorption cross section, $\sigma_\text{abs}$, via $\sum_l N_l {\left|A_l\right|}^2 = \sigma_\text{abs}(\omega)\, \omega^2/\pi$. 
In the expressions above we assumed the geometrical optics limit, 
$\sigma_\text{abs} = 27\pi\, {(\GN M)}^2$, 
known as DeWitt approximation~\cite{DeWitt:1975ys}.
Taking deviations from the approximation at small frequencies into account, a numerical study \cite{Elster:1983pk} found 
$\der E/\der t
\approx \num{7.44e-5}\, \hbar\,{(\GN M)}^{-2}$ such that 
the relative deviation to the DeWitt geometrical optics limit lies at about $6\%$. 

\end{document}